\documentclass[journal,a4paper]{IEEEtran}
\pdfoutput=1

\usepackage{cite}
\usepackage[pdftex]{graphicx}
\usepackage[cmex10]{amsmath}
\usepackage{url}
\usepackage[caption=false,font=footnotesize]{subfig}

\usepackage[T1]{fontenc}
\usepackage[utf8]{inputenc}

\usepackage{rmsbrt_math}
\usepackage{rmsbrt_IBC}
\usepackage{rmsbrt_algorithm}
\usepackage{rmsbrt_sets}
\usepackage{rmsbrt_colours}

\usepackage{dsfont}
\algtext*{EndWhile}
\algtext*{EndIf}
\algtext*{EndFor}

\newcommand*{\Si}{\setS(i)}
\newcommand*{\Sia}{\setS_{a}(i)}
\newcommand*{\Siap}{\setS_{\bar{a}}(i)}

\begin{document}

\title{Globally Optimal Base Station Clustering in Interference Alignment-Based Multicell Networks}

\author{Rasmus~Brandt,~\IEEEmembership{Student~Member,~IEEE,}
        Rami~Mochaourab,~\IEEEmembership{Member,~IEEE,}
        and~Mats~Bengtsson,~\IEEEmembership{Senior~Member,~IEEE}%
\thanks{The authors are with the Department of Signal Processing, ACCESS Linn\ae{}us Centre, School of Electrical Engineering, KTH Royal Institute of Technology, Stockholm, Sweden. E-mails: \texttt{rabr5411@kth.se}, \texttt{ramimo@kth.se}, \texttt{mats.bengtsson@ee.kth.se}.}}

\markboth{Accepted in IEEE Signal Processing Letters}{Brandt \MakeLowercase{\text
it{et al.}}: Globally Optimal Base Station Clustering in Interference Alignment-Based Multicell Networks}

\maketitle

\begin{abstract}
Coordinated precoding based on interference alignment is a promising technique for improving the throughputs in future wireless multicell networks. In small networks, all base stations can typically jointly coordinate their precoding. In large networks however, base station clustering is necessary due to the otherwise overwhelmingly high channel state information (CSI) acquisition overhead. In this work, we provide a branch and bound algorithm for finding the globally optimal base station clustering. The algorithm is mainly intended for benchmarking existing suboptimal clustering schemes. We propose a general model for the user throughputs, which only depends on the long-term CSI statistics. The model assumes intracluster interference alignment and is able to account for the CSI acquisition overhead. By enumerating a search tree using a best-first search and pruning sub-trees in which the optimal solution provably cannot be, the proposed method converges to the optimal solution. The pruning is done using specifically derived bounds, which exploit some assumed structure in the throughput model. It is empirically shown that the proposed method has an average complexity which is orders of magnitude lower than that of exhaustive search.
\end{abstract}

\vspace{-3ex}
\section{Introduction}
For coordinated precoding \cite{OptResAllCoordMultiCellSys} in intermediate to large sized multicell networks, base station clustering \cite{Peters2012,Chen2014,Brandt2016bsubmitted} is necessary for reasons including channel state information (CSI) acquisition overhead, backhaul delays and implementation complexity constraints. In frequency-division duplex mode, the CSI acquisition overhead is due to the feedback required \cite{ElAyach2012,Bolcskei2009}, whereas in time-division duplex mode, the CSI acquisition overhead is due to pilot contamination and allocation \cite{Jose2011b,Mochaourab2015arxiv}.

For the case of interference alignment (IA) precoding \cite{Cadambe2008}, suboptimal base station clustering algorithms have earlier been proposed in \cite{Peters2012} where clusters are orthogonalized and a heuristic algorithm for the grouping was proposed, in \cite{Chen2014} where the clusters are non-orthogonal and a heuristic algorithm on an interference graph was proposed, and in \cite{Brandt2016bsubmitted} where coalition formation and game theory was applied to a generalized frame structure. To the best of the authors' knowledge however, no works in the literature have addressed the problem of finding the globally optimal base station clustering for \mbox{IA-based} systems. Naive exhaustive search over all possible clusterings is not tractable, due to its super-exponential complexity. Yet, the globally optimal base station clustering is important in order to benchmark the more practical schemes in e.g \cite{Peters2012,Chen2014,Brandt2016bsubmitted}. Therefore, in this paper, we propose a structured method based on branch and bound \cite{Land1960,GlobalOptimizationDeterministicApproaches} for finding the globally optimal base station clustering. We consider a generalized throughput model which encompasses the models in \cite{Peters2012,Chen2014,Brandt2016bsubmitted}. When evaluated using the throughput model of \cite{Brandt2016bsubmitted}, empirical evidence shows that the resulting algorithm finds the global optimum at an average complexity which is orders of magnitude lower than that of exhaustive search.

\vspace{-2ex}
\section{Problem Formulation}
We consider a symmetric multicell network where $I$ base stations (BSs) each serve $K$ mobile stations (MSs) in the downlink. A BS together with its served MSs is called a \emph{cell} and we denote the $k$th served MS by BS $i$ as $i_k$. The BSs each have $M$ antennas and the MSs each have $N$ antennas. Each MS is served $d$ spatial data streams. BS $i$ allocates\footnote{Any \emph{fixed} power levels can be used, e.g. obtained from some single-cell power allocation method \cite[Ch.~1.2]{OptResAllCoordMultiCellSys}. Generalizing to \emph{adaptive} multicell power allocation would however lead to loss of tractability in the SINR bound of Thm.~\ref{thm:throughput_bound}, due to $\rho_{i_k}$ not being supermodular \cite{SubmodularFunctionMinimization} when the powers are adaptive.} a power of $P_{i_k}$ to MS $i_k$, in total using a power of $P_i = \sum_{k=1}^K P_{i_k}$, and MS $i_k$ has a thermal noise power of $\sigma_{i_k}^2$. The average large scale fading between BS $j$ and MS $i_k$ is $\gamma_{i_kj}$.

The cooperation between the BSs is determined by the BS clustering, which mathematically is described as a set partition:
\begin{definition}[Set partition] \label{def:set_partition}
    A set partition $\setS = \{ \setC_1, \ldots, \setC_S \}$ is a partition of $\setI = \{ 1, \ldots, I \}$ into disjoint and non-empty sets called clusters, such that $\setC_s \subseteq \setI$ for all $\setC_s \in \setS$ and $\bigcup_{s = 1}^S \setC_s = \setI$. For a cell $i \in \setC_s$, we let $\Si = \setC_s$.
\end{definition}
We assume that IA is used to completely cancel the interference within each cluster.\footnote{Within each cluster, both intra-cell and inter-cell interference is cancelled.} Thus only the intercluster interference remains, which is reflected in the long-term signal-to-interference-and-noise ratios (SINRs) of the MSs:
\begin{assumption}[Signal-to-interference-and-noise ratio]
    Let \\$\rho_{i_k} \!: 2^\setI \rightarrow \realnumbers_+$ be the long-term SINR of MS $i_k$ defined as
    \begin{equation} \label{eq:rho}
        \rho_{i_k}(\Si) = \frac{\gamma_{i_ki} P_{i_k}}{\sigma_{i_k}^2 + \sum_{j \in \setI \setminus \Si} \gamma_{i_kj} P_j}.
    \end{equation}
\end{assumption}
We consider a general model for the MS throughputs, which depends on the cluster size and the long-term SINR. The cluster size determines the overhead, whereas the long-term SINR determines the achievable rate.
\begin{assumption}[Throughput] \label{ass:throughput}
    For a cluster size~$\card{\Si}$ and a long-term SINR~$\rho_{i_k}(\Si)$, the throughput of MS $i_k$ is given by $t_{i_k}(\setS) = v_{i_k}(\card{\Si}, \rho_{i_k}(\Si))$, where $v_{i_k} \!: \naturalnumbers \times \realnumbers_+ \rightarrow \realnumbers_+$ is unimodal in its first argument and non-decreasing in its second argument.
\end{assumption}
The structure of $\rho_{i_k}(\cdot)$ and the monotonicity properties of $v_{i_k}(\cdot,\cdot)$ will be used in the throughput bound to be derived below. The model in Assumption~\ref{ass:throughput} is quite general and is compatible with several existing throughput models:

\begin{example}
    In \cite{Peters2012}, the clusters are orthogonalized using time sharing, and no intercluster interference is thus received. A coherence time of $L_c$ is available. Each BS owns $1/I$ of the coherence time, which is contributed to the corresponding cluster. Larger clusters give more time for data transmission but also require more CSI feedback, which in \cite{Peters2012} is modelled as a quadratic function, giving the throughput model as:
    \begin{equation} \label{eq:example_Peters2012}
        v_{i_k}(\card{\Si}, \cdot) = \left( \frac{\card{\Si}}{I} - \frac{\card{\Si}^2}{L_c} \right) d \log \left( 1 + \varrho_{i_k} \right).
    \end{equation}
    where $\varrho_{i_k} = \frac{\gamma_{i_ki} P_{i_k}}{\sigma_{i_k}^2}$ is the constant \emph{signal-to-noise ratio} (SNR). The function in \eqref{eq:example_Peters2012} is strictly unimodal in its first argument and independent of its second argument.
\end{example}
\begin{example}
    In \cite{Chen2014}, the clusters are operating using spectrum sharing. The CSI acquisition overhead is not accounted for. A slightly modified\footnote{The original SINR model in \cite{Chen2014} includes the impact of the instantaneous IA filters, which we neglect here in order to avoid the cross-dependence between the IA solution and the clustering. This corresponds to how the approximated interference graph weights are derived in \cite{Chen2014}.} version of their throughput model is then: 
    \begin{equation} \label{eq:example_Chen2014}
        v_{i_k}(\cdot, \rho_{i_k}(\Si)) = d \log \left( 1 + \rho_{i_k}(\Si) \right).
    \end{equation}
    The function in \eqref{eq:example_Chen2014} is independent of its first argument, and strictly increasing in its second argument.
\end{example}
\begin{example}
    In \cite{Brandt2016bsubmitted}, intercluster time sharing and intercluster spectrum sharing are used in two different orthogonal phases. For the CSI acquisition overhead model during the time sharing phase, a model similar to the one in \cite{Peters2012} is used. For the achievable rates during the spectrum sharing phase, \mbox{long-term} averages are derived involving an exponential integral. The model is thus
    \begin{equation} \label{eq:example_Brandt2016b}
        v_{i_k}(\card{\Si}, \rho_{i_k}(\Si)) = \alpha_{i_k}^{(1)}(\card{\Si}) \, r_{i_k}^{(1)} + r_{i_k}^{(2)}(\rho_{i_k}(\Si))
    \end{equation}
    where
    \begin{align*}
        &\alpha_{i_k}^{(1)}(\card{\Si}) = \frac{\card{\Si}}{I} - \frac{(M + K(N+d)) \card{\Si} + KM \card{\Si}^2}{L_c}, \\
        &r_{i_k}^{(1)} = d \, e^{1/\varrho_{i_k}} \int_{1/\varrho_{i_k}}^\infty t^{-1} e^{-t} \, \mathrm{d}t, \\
        &r_{i_k}^{(2)}(\rho_{i_k}(\Si)) = d \, e^{1/\rho_{i_k}(\Si)} \int_{1/\rho_{i_k}(\Si)}^\infty t^{-1} e^{-t} \, \mathrm{d}t.
    \end{align*}
    The function in \eqref{eq:example_Brandt2016b} is strictly unimodal in its first argument and strictly increasing in its second argument.
\end{example}

Given the MS throughput model, we introduce the notion of a system-level objective:
\begin{definition}[Objective] \label{def:objective}
    The performance of the entire multicell system is given by $f(\setS)~=~g(t_{1_1}(\setS), \ldots, t_{I_K}(\setS))$, where $g \!: \realnumbers_+^{I \cdot K} \rightarrow \realnumbers_+$ is an argument-wise non-decreasing function. 
\end{definition}
The function $f(\setS)$ thus maps a set partition to the corresponding system-level objective. Typical examples of objective functions are the weighted sum $f_\text{WSR}(\setS)~=~\sum_{(i,k)} \lambda_{i_k} t_{i_k}(\setS)$ and the minimum weighted throughput $f_\text{min}(\setS)~=~\min_{(i,k)} \lambda_{i_k} t_{i_k}(\setS)$.

\section{Globally Optimal Base Station Clustering} \label{sec:branchandbound}
We will now provide a method for solving the following combinatorial optimization problem:
\begin{equation} \label{opt:system}
    \begin{aligned}
    \setS^\star = \, & \underset{\setS}{\text{arg\,max}}
    & & f(\setS) \\
    & \text{subject to}
    & & \text{$\setS$ satisfying Def.~\ref{def:set_partition}} \\
    & & & \card{\Si} \leq D, \; \forall \, i \in \setI.
    \end{aligned}
\end{equation}
The cardinality constraint is used to model cluster size constraints due to IA feasibility \cite{Liu2013}, CSI acquisition feasibility \cite{Brandt2016bsubmitted}, implementation feasibility, etc.

\subsection{Restricted Growth Strings and Exhaustive Search}
In the algorithm to be proposed, we use the following alternate representation of a set partition:
\begin{definition}[Restricted growth string, {\cite[Sec.~7.2.1.5]{TAoCPVol4Apart1}}]
    A set partition $\setS$ can equivalently be expressed using a \emph{restricted growth string} $a = a_1a_2\ldots a_I$ with the property that $a_i~\leq~1~+~\max(a_1, \ldots, a_{i-1})$ for $i \in \setI$. Then $a_i \in \naturalnumbers$ describes which cluster that cell $i$ belongs to. We let $\setS_a$ denote the mapping from $a$ to the set partition $\setS$, and $a_\setS$ as its inverse.
\end{definition}
For example, the set partition $\setS_a~=~\{ \{1, 3 \}, \{ 2 \} , \{ 4 \} \}$ would be encoded as $a_\setS~=~1213$. One approach to solving the optimization problem in \eqref{opt:system} is now by enumerating all restricted growth strings of length $I$, using e.g. Alg.~H of \cite[Sec.~7.2.1.5]{TAoCPVol4Apart1}. The complexity of this approach is however $\mathbb{B}_I$, the $I$th \emph{Bell number}\footnote{The $I$th Bell number describes the number of set partitions of $\setI$ \cite[p. 287]{IntroductoryCombinatorics}, and can be bounded as $\mathbb{B}_I~<~\left( 0.792I/\log(1 + I) \right)^I$ \cite{Berend2010}. The 17 first Bell numbers are 1, 1, 2, 5, 15, 52, 203, 877, 4\,140, 21\,147, 115\,975, 678\,570, 4\,213\,597, 27\,644\,437, 190\,899\,322, 1\,382\,958\,545, 10\,480\,142\,147.}, which grows super-exponentially.

\subsection{Branch and Bound Algorithm}
\begin{figure}[t]
    \centering
    \includegraphics[width=\columnwidth]{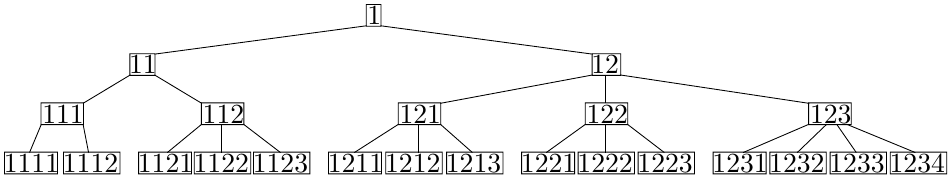}
    \caption{Example of branch and bound search tree for $I = 4$.} \label{fig:searchtree}
\end{figure}

Most of the possible set partitions are typically not interesting in the sense of the objective of \eqref{opt:system}. For example, most set partitions will include clusters whose members are placed far apart, thus leading to low SINRs. By prioritizing set partitions with a potential to achieve large throughputs, the complexity of finding the globally optimal set partition can be decreased significantly compared to that of exhaustive search. This is the idea of the branch and bound approach \cite{Land1960}, which entails bounding the optimal value $f(\setS^\star)$ from above and below for a sequence of \emph{partial solutions}. When the bounds converge, the optimal solution has been found. The partial solutions are described using \emph{partial restricted growth strings}:
\begin{definition}[Partial restricted growth string]
    The restricted growth string $\bar{a} = \bar{a}_1\bar{a}_2\ldots \bar{a}_l$ is \emph{partial} if $l = \textsc{length}(\bar{a}) \leq I$. The corresponding partial set partition, where only the first $l$ cells are constrained into clusters, is denoted $\setS_{\bar{a}}$.
\end{definition}
The branch and bound method considers the sequence of partial solutions by dynamically exploring a search tree (see Fig.~\ref{fig:searchtree}, at the top of the page), in which each interior node corresponds to a partial restricted growth string. By starting at the root and traversing down the search tree\footnote{At level $i \leq I$ of the tree, there are $\mathbb{B}_i$ nodes.}, more cells are constrained into clusters, ultimately giving the leaves of the tree which describe all possible restricted growth strings.

\subsubsection{Bounds}
We now provide the bounds that will be used to avoid exploring large parts of the search tree.
\begin{lemma}[Objective bound]
    Let $\check{t}_{i_k}(\setS_{\bar{a}})$ be an upper bound of the throughput of MS $i_k$ for all leaf nodes in the sub-tree below the node described by $\bar{a}$. Then the function 
    \begin{equation*}
        \check{f}(\setS_{\bar{a}})~=~g(\check{t}_{1_1}(\setS_{\bar{a}}), \ldots, \check{t}_{I_K}(\setS_{\bar{a}}))
    \end{equation*}
    is an upper bound of the objective in \eqref{opt:system} for all leaves in the sub-tree below the node described by $\bar{a}$.
\end{lemma}
\begin{IEEEproof}
    This follows directly from the argument-wise monotonicity of $g(t_{1_1}, \ldots, t_{I_K})$ in Def.~\ref{def:objective}.
\end{IEEEproof}

In order to describe the throughput bound, we will introduce three sets. Given a node described by $\bar{a}$, the cells in $\setP_{\bar{a}} = \{ 1, \ldots, \textsc{length}(\bar{a}) \} \subseteq \setI$ are constrained into clusters as given by $\setS_{\bar{a}}$. The remaining cells in $\setP^\bot_{\bar{a}} = \setI \setminus \setP_{\bar{a}}$ are still unconstrained.\footnote{In the sub-tree below the node described by $\bar{a}$, there is a leaf node for all possible ways of constraining the cells in $\setP^\bot_{\bar{a}}$ into clusters.} The set of cells which could accommodate more members in the corresponding clusters\footnote{For the sake of this definition, we consider the non-constrained cells in $\setP^\bot_{\bar{a}}$ to be in singleton clusters.} are written as $\setF_{\bar{a}} = \left\{ i \in \setP_{\bar{a}} : \card{\Siap} < D \right\} \cup \setP^\bot_{\bar{a}}$.
\begin{theorem}[Throughput bound] \label{thm:throughput_bound}
    Let $t_{i_k}(\setS_a)$ be the throughput of MS $i_k$ for some leaf node in the sub-tree below the node described by $\bar{a}$. It can be bounded as $t_{i_k}(\setS_a)~=~v_{i_k}(\card{\Sia}, \rho_{i_k}(\Sia)) \leq v_{i_k}(\check{B}_{i_k}, \check{\rho}_{i_k})$ where
    \begin{align*}
        \check{B}_{i_k} &=
        \begin{cases}
            \card{\Siap} & \textnormal{if} \; \card{\Siap} \geq B_{i_k}^\star, \\
            \min \left( \card{\Siap} + \card{\setP^\bot_{\bar{a}}}, B_{i_k}^\star \right) & \textnormal{else if} \; i \in \setP_{\bar{a}}, \\
            \min \left( \card{\setF_{\bar{a}}}, B_{i_k}^\star \right) & \textnormal{else if} \; i \in \setP^\bot_{\bar{a}},
        \end{cases} \\
        B_{i_k}^\star &= \argmax_{b \in \naturalnumbers, b \leq D} t_{i_k}(b, \check{\rho}_{i_k}),
    \end{align*}
    and
    \begin{align}
        \check{\rho}_{i_k} = \, &\underset{\card{\setE_{i_k}} \leq D}{\textnormal{maximize}}
        & & \rho_{i_k}(\setE_{i_k}) \label{opt:rho_bound} \\
        & \textnormal{subject to}
        & & \textnormal{if} \; i \in \setP_{\bar{a}} \textnormal{:} \notag \\
        & & & \hspace{2em} \Siap \subseteq \setE_{i_k} \subseteq \left( \Siap \cup \setP^\bot_{\bar{a}} \right) \notag \\
        & & & \textnormal{else if} \; i \in \setP^\bot_{\bar{a}} \textnormal{:} \notag \\
        & & & \hspace{2em} \setE_{i_k} \subseteq \setF_{\bar{a}}. \notag
    \end{align}
\end{theorem}
\begin{IEEEproof}
    First note that $\check{\rho}_{i_k}$ is an upper bound of the achievable long-term SINR for MS $i_k$ in the considered sub-tree, since the requirement of disjoint clusters is not enforced in the optimization problems\footnote{The optimal solution to the optimization problem in \eqref{opt:rho_bound} can be found by minimizing the denominator of $\rho_{i_k}(\setE_{i_k})$ in \eqref{eq:rho}, which is easily done using greedy search over the feasible set. The set-function $\rho_{i_k}(\setE_{i_k})$ is \emph{supermodular} \cite{SubmodularFunctionMinimization}, i.e. demonstrating ``increasing returns'', which is the structure that admits the simple solution of the optimization problem. Without changes, Thm.~\ref{thm:throughput_bound} would indeed hold for any other supermodular set-function $\rho_{i_k}(\setE_{i_k})$.} in \eqref{opt:rho_bound}. We therefore have that $v_{i_k}(\card{\Sia}, \rho_{i_k}(\Sia)) \leq v_{i_k}(\card{\Sia}, \check{\rho}_{i_k})$, due to the monotonicity property of $v_{i_k}(\cdot,\cdot)$. Now the fact that $v_{i_k}(\card{\Sia}, \check{\rho}_{i_k})~\leq~v_{i_k}(\check{B}_{i_k}, \check{\rho}_{i_k})$ holds is proven. Note that $B_{i_k}^\star$ is the optimal size of the cluster, in terms of the first parameter of $v_{i_k}(\cdot,\check{\rho}_{i_k})$. If $\card{\Siap} \geq B_{i_k}^\star$, the cluster is already larger than what is optimal, and keeping the size is thus a bound for all leaves in the sub-tree. On the other hand, if $\card{\Siap} < B_{i_k}^\star$ and $i \in \setP_{\bar{a}}$, $\check{B}_{i_k}$ is selected as close to $B_{i_k}^\star$ as possible, given the number of unconstrained cells that could conceivably be constrained into $\Siap$ further down in the sub-tree. If $i \in \setP^\bot_{\bar{a}}$ however, we similarly bound $\check{B}_{i_k}$, except that we only consider cells in non-full clusters for cell $i$ to conceivably be constrained to further down in the sub-tree. Due to the unimodality property of $v_{i_k}(\cdot,\check{\rho}_{i_k})$ and the fact that $\check{B}_{i_k}$ is selected optimistically, we have that $v_{i_k}(\card{\Sia}, \check{\rho}_{i_k}) \leq v_{i_k}(\check{B}_{i_k}, \check{\rho}_{i_k})$, which gives the bound.
\end{IEEEproof}
As the algorithm explores nodes deeper in the search tree, $\textsc{length} \left( \bar{a} \right)$ gets closer to $I$, and there is less freedom in the bounds. For $\textsc{length} \left( \bar{a} \right) = I$, the bounds are tight.

\algrenewcommand\algorithmicindent{0.25em}%
\begin{algorithm}[t]
    \caption{Branch and Bound for Base Station Clustering} \label{alg:branchandbound}
    \begin{algorithmic}[1]
        \Require Initial $a_\text{incumbent}$ from some heuristic, $\epsilon \geq 0$
        \State $\texttt{live} \gets [1]$
        \While{$\textsc{length}(\texttt{live}) > 0$}
            \State $\bar{a}_\text{parent} \gets \text{node from \texttt{live} with highest upper bound}$
            \IIf{$\check{f}(\setS_{\bar{a}_\text{parent}}) - f(\setS_{a_\text{incumbent}}) < \epsilon$} \Goto{line \ref{alg:branchandbound:final}} \EndIIf
            \ForAll{$\bar{a}_\text{child}$ from $\textsc{branch}(\bar{a}_\text{parent})$}
                \If{$\check{f}(\setS_{\bar{a}_\text{child}}) > f(\setS_{a_\text{incumbent}})$}
                    \If{$\textsc{length}(\bar{a}_\text{child}) = I$}
                        \State $a_\text{incumbent} \gets \bar{a}_\text{child}$
                    \Else
                        \State Append $\bar{a}_\text{child}$ to \texttt{live}
                    \EndIf
                \EndIf
            \EndFor
        \EndWhile
        \State \Return globally optimal $a_\text{optimal} = a_\text{incumbent}$ \label{alg:branchandbound:final}
    \end{algorithmic}
\end{algorithm}
\algrenewcommand\algorithmicindent{1em}%
\begin{figure}[t]
    \vspace{-1em}
    \hrule
    \vspace{0.5em}
    \begin{algorithmic}[1]
        \Function{branch}{$\bar{a}_\text{parent}$}
            \State Initialize empty list $\texttt{children} = []$
            \For{$b = 1:(1 + \max(\bar{a}_\text{parent}))$}
                \State Append $[\bar{a}_\text{parent}, b]$ to \texttt{children}
            \EndFor
            \State \Return \texttt{children}
        \EndFunction
    \end{algorithmic}
    \vspace{0.5em}
    \hrule
    \vspace{-1em}
\end{figure}

\subsubsection{Algorithm}
The proposed branch and bound method is described in Alg.~\ref{alg:branchandbound}. The algorithm starts by getting an initial incumbent solution from a heuristic (e.g. from Sec.~\ref{sec:heuristic} or \cite{Peters2012,Chen2014,Brandt2016bsubmitted}), and then sequentially studies the sub-tree which currently has the highest upper bound. By comparing the upper bound $\check{f}(\setS_{\bar{a}})$ to the currently best lower bound $f(\setS_{a_\text{incumbent}}) \leq f(\setS^\star)$, the \emph{incumbent solution}, the sub-tree below $\bar{a}$ can be \emph{pruned} if it provably cannot contain the optimal solution, i.e. if $\check{f}(\setS_{\bar{a}})~<~f(\setS_{a_\text{incumbent}})$. If a node $\bar{a}$ cannot be pruned, all children of $\bar{a}$ are built by a branching function and stored in a list for future exploration by the algorithm. If large parts of the search tree can be pruned, few nodes need to be explicitly explored, leading to a complexity reduction. The algorithm ends when the optimality gap for the current incumbent solution is less than a pre-defined $\epsilon \geq 0$.

\begin{theorem}
    Alg.~\ref{alg:branchandbound} converges to an $\epsilon$-optimal solution of the optimization problem in \eqref{opt:system} in at most $\sum_{i = 1}^I \mathbb{B}_i$ iterations.
\end{theorem}
\begin{IEEEproof}
    Only sub-trees in which the optimal solution cannot be are pruned. Since all non-pruned leaves are explored, the global optimum will be found. No more than all $\sum_{i = 1}^I \mathbb{B}_I$ nodes of the search tree can be traversed.
\end{IEEEproof}
In Sec.~\ref{sec:numerical_results} we empirically show that the average complexity is significantly lower than the worst case.

\vspace{-2ex}
\subsection{Heuristic Base Station Clustering} \label{sec:heuristic}
We also provide a heuristic (see Alg.~\ref{alg:heuristic}) which can be used as the initial incumbent in Alg.~\ref{alg:branchandbound}, or as a low complexity clustering algorithm in its own right. The heuristic works by greedily maximizing a function of the average channel gains in the clusters while respecting the cluster size constraint. The heuristic is similar to Ward's method \cite{Ward1963}.

\begin{algorithm}[t]
    \caption{Heuristic for Base Station Clustering} \label{alg:heuristic}
    \begin{algorithmic}[1]
        \Require $\setL = \{ (i,j) \in \setI \times \setI \mid i \neq j \}$, $\setS = \{ \{ 1 \}, \ldots, \{ I \} \}$
        \While{$\card{\setL} > 0$}
            \State $(i^\star, j^\star) = \argmax_{(i,j) \in \setL} \sum_{k=1}^K \log \left( 1 + \gamma_{i_kj} P_j/\sigma_{i_k}^2 \right)$
            \State Let $\Xi_{\left( i^\star, j^\star \right)} \gets \setS(i^\star) \cup \setS(j^\star)$
            \If{$\card{\Xi_{\left( i^\star, j^\star \right)}} \leq D$}
                \State Let $\setS \gets \left( \setS \setminus \{ \setS(i^\star), \setS(j^\star) \} \right) \cup \{ \Xi_{\left( i^\star, j^\star \right)} \}$
            \EndIf
            \State Let $\setL \gets \setL \setminus \{ \left( i^\star, j^\star \right) \}$
        \EndWhile
        \State \Return heuristic solution $a_\text{heuristic} = a_\setS$
    \end{algorithmic}
\end{algorithm}

\begin{figure}[t]
    \centering
    \includegraphics{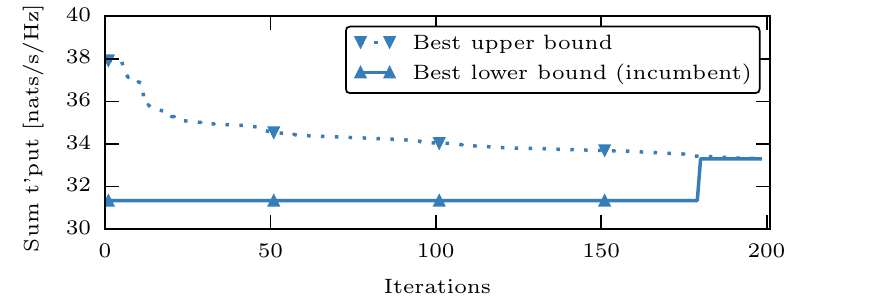}
    \vspace{-5ex}
    \caption{Example of convergence of the algorithm for one realization.} \label{fig:convergence_bounds}
    \vspace{2ex}
    \includegraphics{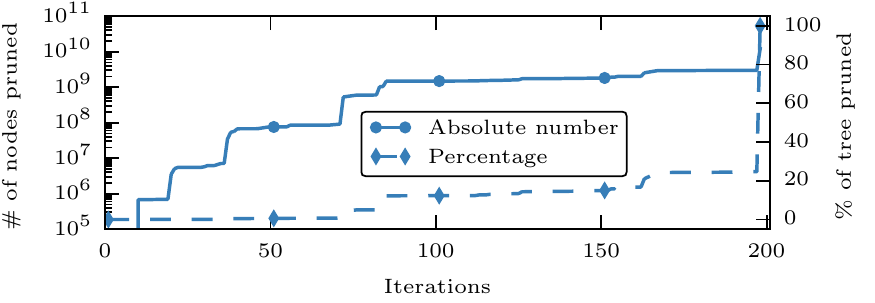}
    \vspace{-5ex}
    \caption{Pruning evolution for the realization in Fig.~\ref{fig:convergence_bounds}.} \label{fig:convergence_fathom}
    \vspace{2ex}
    \includegraphics{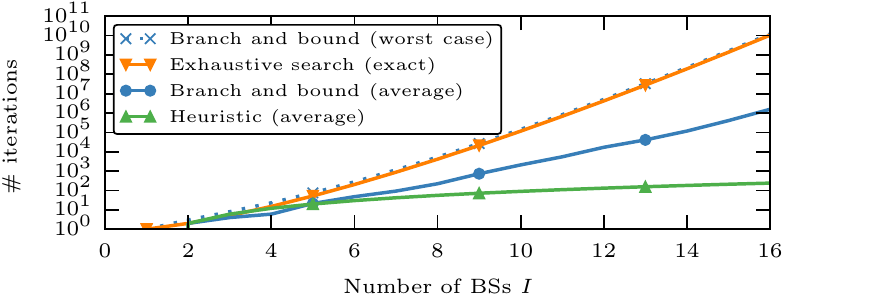}
    \vspace{-5ex}
    \caption{Average complexity as a function of $I$.} \label{fig:I}
    \vspace{2ex}
    \includegraphics{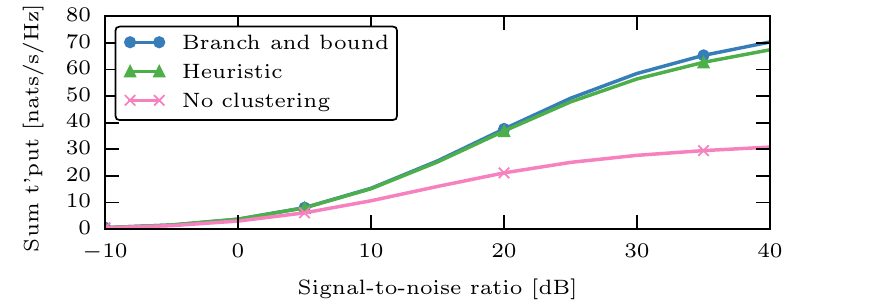}
    \vspace{-5ex}
    \caption{Sum throughput performance as a function of \texttt{SNR}.} \label{fig:SNR}
\end{figure}

\vspace{-2ex}
\section{Numerical Results} \label{sec:numerical_results}
For the performance evaluation \cite{Brandt2015c_code}, we consider a network of $I = 16$ BSs, $K = 2$ MSs per cell, and $d = 1$ per MS. We employ the throughput model from \eqref{eq:example_Brandt2016b} and let $f(\setS)~=~\sum_{(i,k)} t_{i_k}(\setS)$. We let the number of antennas be $M = 8$ and $N = 2$. This gives a hard size constraint as $D = 4$ cells per cluster, due to IA feasibility \cite{Liu2013}. We consider a large-scale setting with path loss $15.3 + 37.6 \log_{10}(\text{distance} \, \text{[m]})$, i.i.d. log-normal shadow fading with $8$ dB std. dev., and i.i.d. $\mathcal{CN}(0,1)$ small-scale fading. The BSs are randomly dropped in a $2000 \times 2000 \, \text{m}^2$ square and the BS-MS distance is $250 \, \text{m}$. We let $L_c = 2\,700$, corresponding to an MS speed of $30$ km/h at a typical carrier frequency and coherence bandwidth \cite{Jindal2010}.

In Fig.~\ref{fig:convergence_bounds} we show the convergence of the best upper bound and the incumbent solution, respectively, for one network realization with $\texttt{SNR} = P_{i_k}/\sigma_{i_k}^2 = 20 \, \text{dB}$. The number of iterations needed was $198$ and a total of $908$ nodes were bounded. Naive exhaustive search would have needed exploring $\mathbb{B}_{16}~=~10\,480\,109\,379$ nodes, and the proposed algorithm was thus around $1 \cdot 10^7$ times more efficient for this realization\footnote{Also note that $\sum_{i=1}^{16} \mathbb{B}_i = 12\,086\,679\,035$, i.e. the actual running time of the algorithm was significantly lower than the worst-case running time.}. The number and fraction of nodes pruned during the iterations is shown in Fig.~\ref{fig:convergence_fathom}. At convergence, $99.99999 \%$ of the search tree had been pruned.

We show the average number of iterations as a function of network size in Fig.~\ref{fig:I}. The complexity of the proposed algorithm is orders of magnitude lower than the complexity of exhaustive search.

In Fig.~\ref{fig:SNR} we show the sum throughput performance as a function of $\texttt{SNR}$, averaged over $250$ network realizations. The heuristic algorithm performs well: it is close to the optimum\footnote{For this network size, exhaustive search is not tractable. Without Alg.~\ref{alg:branchandbound}, we would thus not know that Alg.~\ref{alg:heuristic} performs so well. This shows the significance of Alg.~\ref{alg:branchandbound} as a benchmarking tool for practical but suboptimal clustering algorithms such as Alg.~\ref{alg:heuristic}, or the algorithms in \cite{Peters2012,Chen2014,Brandt2016bsubmitted}.}, and has about twice the throughput of the no clustering case, where $\setS = \{ \{ 1 \}, \ldots, \{ I \} \}$. The grand cluster $\setS = \{ \setI \}$ has zero sum throughput since $I > D$, and is therefore not shown.

\section{Conclusions}
With a structured branch and bound approach, the otherwise intractable base station clustering problem has been solved. The algorithm is intended for benchmarking of suboptimal base station clustering heuristics in intermediate size networks.

\vfill\pagebreak

\bibliographystyle{IEEEtran}
\bibliography{IEEEabrv,coordinated_precoding,rasmus_brandt}

\end{document}